\documentclass[12pt,oneside,a4paper]{article}
\usepackage{hyperref}
\usepackage{graphicx}
\begin{document}
\title{Final fate of Kantowski-Sachs gravitational collapse}
  \author{B. Terez\'on \\  \href{mailto:{\it brisa.terezon@udb.edu.sv}}{\it{brisa.terezon@udb.edu.sv}}    \\ M. de Campos \\ 
   \href{mailto:miguel.campos@ufrr.br}{\it{miguel.campos@ufrr.br} }}
 
\maketitle
%
%
\begin{abstract}
Although it is not a fundamental question, to determine exact and general solutions for a given theory has advantages over a numerical integration in many specific cases. Of course, respecting the peculiarities of the problem. Revisiting the integration of the General Relativity Theory field equations for the Kantowski-Sachs spacetime, that describes a homogeneous but anisotropic universe whose spatial section has the topology of $R \times S^2$, we integrate the equations for arbitrary curvature parameter, and writing the solutions considering the process of gravitational collapse.  We took the opportunity and made some comments involving some features of the model such as: energy density, shear, viscosity, and the production of gravitational waves via Petrov classification.
\end{abstract}
\section{Introduction}	
Presently, we have, in the scientific literature, some solutions for the  general relativity theory field equations, but this does not means that the integration process of these equations is a simple task.  Especially, if we evade from the usual homogeneous and isotropic framework.  

The Kantowski-Sachs model (we'll simplify as KS model from here on out) has returned to the scientific interest, and to situate the reader in what 
has been published about it, let's start by the study of Mendez and Henriquez \cite{Mendez}, that study an
inflationary era for the KS model.  On the other hand, Chakraborty and Roy \cite{Roy} deal with an anisotropic
cosmological model considering the gravitational constant plus a $\Lambda$-term with temporal dependence.  
A little later, Adhav and other authors \cite{Adhav} also considered an anisotropic model, now, to include 
a dark energy component in a discussion of an universal dynamics, and a study involving a skyrme fluid source presents several interesting quotes from works done
within KS geometry \cite{Skyrme}.  Still from the cosmological point of view  we have the work of Katore and Hatkar \cite{Katore} 
that deals with the KS model in $f(R,T)$ theory. We also have the construction of cosmological scenarios within the $f(R)$ theory, focused on KS geometry, where an approach draws attention to the formation of a bounce and turnaround, as well as cyclic cosmology \cite{Saridarkis}.  More recently, appear \cite{Ray} a work where the authors constructed an anisotropic a cosmological model with a  dark energy component and a magnetized fluid, looking to keep the advantages of a nontrivial isotropization history due to the presence of an anisotropic energy source, in spite of the inflation.  

The cosmological point of view and the gravitational collapse process can be considered as the two sides of the same coin, and turning to the other side, we have the work of Abbas \cite{Abbas} that consider the gravitational collapse in the presence of a magnetic field; while Maurya,  and other authors, study anisotropic models for compact stars \cite{Maurya}.  A little apart from these lines, we can also mention the work of Bradley and others authors \cite{Bradley} involving scalar perturbations.  Although, distancing itself  a little from the KS geometry, we could not fail to mention Doroshkevich's article on the universe model with an uniform magnetic field \cite{Dorosh}.

We restrict ourselves to the Einstein's theory framework, and summarizing the opera:  
we will do a study of the collapsing process analysis involving it's final destination, considering a Kantowski-Sachs spacetime; write the Kretschmann and Weyl scalars and estimate the possibility of gravitational waves production using Petrov classification.  We analyze the adoption of some ansatz considered in the literature and their viability, confronting with our exact solutions, and look for  different possibilities of partitioning the velocity gradient of an element of the fluid (naturally linked to the scalar shear of the model), suggesting some perspective for future work.

On the other hand, also recently, the detection of gravitational waves \cite{Abbo} generated by the merger of two relatively light black holes, with 7$\odot$ and 12$\odot$,  that occurred about a billion of light-years from us, provides a vigorous increase in this particular topic of scientific research interest. We can not forget to remember that the gravitational waves were first predicted theoretically by Einstein himself, still at the time of the emergence of the general relativity theory. As we are dealing with the collapsing process, even more being of anisotropic character, it is expected that in this case we have the propagation of gravitational waves, and that these have a signature correlated with the characteristics of the model of collapse that we studied in this work.
\section{Basic equations of the model}\label{sec:1}
In fact, when we speak of Kantowski-Sachs model this is characterized by have a positive curvature.
However, we will adopt a similar procedure to others authors who lend to the curvature of KS model an arbitrary feature, at first, seeking
a generalization of the original model \cite{Kantowski}.  Doing so, we consider the spacetime governed by the metric \cite{Ellis}
\begin{equation}\label{metric}
ds^2=-dt^2+A(t)^2dr^2+B(t)^2d\theta^2+B(t)^2f(\theta)^2d\phi^2\, , 
\end{equation}
where $A(t)$ and $B(t)$ are functions only of time, and the function $f(\theta)$, only of the coordinate $\theta$.
The function $f(\theta)$ can assume the following dependences:
\begin{equation}\label{f(y)}
f(\theta)=
\left( \begin{array}{c}
	\sin \theta  \\ 
	 \theta  \\
	 \sinh \theta
\end{array} \right)
\end{equation}
and it will be clear observing the field equations the connection between $f(\theta)$ and the spacetime curvature.

For the energy momentum tensor, we adopt the  usual perfect fluid form
\begin{equation}\label{momentum}
T_{\mu\nu}=(\rho+p)u_{\mu}u_{\nu}+pg_{\mu\nu},
\end{equation}
where $\rho$ is the energy density, $P$ is the thermodynamical pressure, and $u$ is the four-velocity of the fluid.

Hence, the field equations for the general relativity theory, namely
\begin{equation}
G_{\mu \nu} = \kappa T_{\mu \nu }\ \, , \, \kappa = 8\pi G/c^2 \, ,
\end{equation}
assumes the following form
\begin{eqnarray}
\frac{2\dot{A}\dot{B}}{AB}-\left(\frac{\partial ^2 f/{\partial \theta^2}}{f}\right)\frac{1}{B^{2}}+\frac{\dot{B}^{2}}{B^{2}}&=& \kappa\rho, \label{fe1}\\
\frac{2\ddot{B}}{B}+\left(\frac{\partial ^2 f/{\partial \theta^2}}{f}\right)\frac{1}{B^{2}}+\frac{\dot{B}^{2}}{B^{2}}&=& -\kappa p_1 \label{fe2}\\
\frac{\ddot{B}}{B}+\frac{\dot{A}\dot{B}}{AB}+\frac{\ddot{A}}{A}&=& -\kappa p_2,\label{fe3}
\end{eqnarray}
where the dot means $\dot{B} = \partial B /\partial t$.

Although we may consider distinct pressures for different components of the energy momentum tensor, by virtue of our fluid 
being perfect, $P_1 = P_2$.  The inclusion of the pressure, especially in the gravitational collapse process, increases the difficulty
of field equations integration, and also the precise determination of some physical concepts, as the mass and volume.  An identical note can be made in the case of inclusion of a vacuum energy term, $\Lambda (t)$, that also generates a pressure in the field equations, and consequently to a reinterpretation of the energy momentum tensor.

So, at a first approach, we consider a dust fluid, and the integration of Eq.(\ref{fe2}) results:
\begin{equation}\label{tet}
t=c_{2}\pm\frac{\sqrt{c_{1}B-KB^{2} }}{K}\mp\frac{c_{1}\tan^{-1}\left\{K^{1/2}\frac{[B-c_{1}/2K]}{\sqrt{c_{1}B-KB^{2} }}\right\}}{2K^{3/2}} \, ,
\end{equation}
where $c_1$ and $c_2$ are integration constants, and $K $ is the curvature ( defined by $K = -\frac{\partial ^2 f}{\partial \theta^2}/f$).  Following the definition and the Eq.(\ref{f(y)}), 
we have $K = +1,\, 0, \, -1$.

However, an explicit solution would be welcome, given that we still have an additional scale factor to be determined.
Hence, introducing the variable   $Z = \sqrt{B}$, Eq.(\ref{fe2}) assumes the form
\begin{equation}\label{free}
Z^{\prime \prime}+\frac{K}{4}Z = 0\, .
\end{equation}
The Eq.(\ref{free}) is similar to free particle for $K=0$, and for $K=+1$ to the simple harmonic motion.  Consequently, a general solution for Eq.(\ref{free}) can be written as:
\begin{equation}\label{B(eta)}
B(\eta) = \frac{1}{K}\left \lbrace C_1\sin \frac{\sqrt{K}}{2}\eta+ C_2\cos \frac{\sqrt{K}}{2}\eta \right \rbrace ^2 \, ,
\end{equation}
where $dt = B d\eta$, and $B^\prime = \partial B/ \partial \eta$.

Taking into account the conditions
\begin{eqnarray}
\eta = \eta_{BC} &\rightarrow& B=0  \nonumber \\ 
\eta =0 &\rightarrow & B=B_i \nonumber \, ,
\end{eqnarray}
and proper integration constants, Eq.(\ref{B(eta)}) assumes the form:
\begin{equation}\label{geral}
B(\eta)=B_{\ast}\left\{\sqrt{1/K}\sin\left[\frac{\sqrt{K}}{2}\left(\eta_{BC}-\eta\right)\right]\right\}^{2}\, ,
\end{equation}
where $B_{\ast} = \frac{\sqrt{K}B_i}{\sin (\sqrt{K} \eta_{BC}/2)}$.  We have that $B_i$ is the value for the scale factor $B(\eta)$ at the beginning of the collapsing process (that eventually can be considered unity);  and $\eta_{BC}$ is  the {\it conformal} time necessary for the scale factor $B(\eta)$ collapse.  It turns out that the collapsing time is closely linked to the curvature spacetime, and may can decide the fate of the star fluid in respect to the singularity 
be naked, or if we have the formation of a black object \cite{Campos,Perico}.

Turning to the field equations, let's look for the for $A(\eta)$ scale factor expression.  Rewritten Eq.(\ref{fe3}) in terms of conformal time,
we obtain:
\begin{equation}\label{conformal}
\frac{B^{\prime \prime }}{B}+\frac{A^{\prime \prime }}{A}-(\frac{B^\prime}{B})^2 = 0 \,,
\end{equation}
and substituting Eq.(\ref{geral}), integration furnishes
\begin{eqnarray}\label{A-eta}
A(\eta)= &C_{1}&\frac{1+\tan (\sqrt{K}\eta /2)}{\tan (\sqrt{K}\eta /2)-1} + \\ \nonumber
 &C_2&\left \lbrace 2+\sqrt{K}\eta /2 \tan (\sqrt{K}\eta /2) + \sqrt{K}\eta /2  \right\rbrace/(\tan (\sqrt{K} \eta /2 -1))\, ,
\end{eqnarray}
where  $C_1$ and $C_2$ are new integration constants.

It seemed to us, therefore, more reasonable to consider each case of curvature separately, mainly due to the mixing of the initial conditions of both scale factors.  The nesting of initial conditions of the kind that happens here is possible to be circumvent as long as we explicitly have the relationship, in a given moment, between the conditions of both scale factors.   
So, proceeding like this: \\
$\bullet$  {\it \underline{Case with null curvature}}\\
In this case the Eq.(\ref{geral}) and the Eq.(\ref{A-eta}) assume the following forms
\begin{eqnarray}
B(\eta) &=& B_i \left\lbrace  1-\eta/\eta_{BC}   \right\rbrace    ^2  \\
A(\eta) &=& A_i  \left\lbrace  [  1-\eta/\eta_{BC}]        ^2 + \frac{1}{1-\eta/\eta_{BC}} \right\rbrace \, ,
\end{eqnarray}   
respectively.  

\begin{figure}[pb]
\begin{center}
	\includegraphics[scale=0.7]{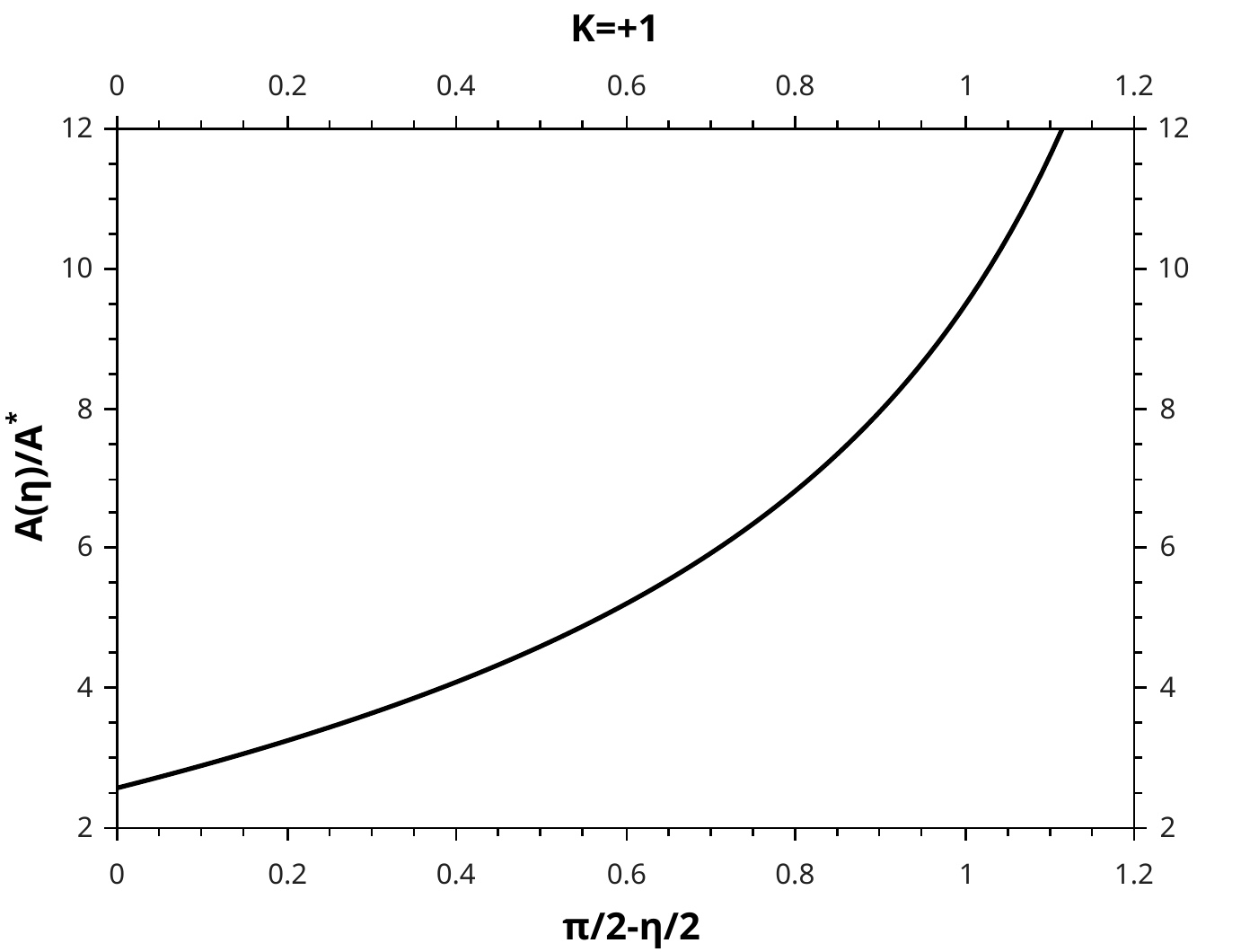}
	\caption{ Evolution for the scale factor $B(\eta)$ and $A(\eta)$ for a null curvature.
				We  have the collapses of $B(\eta)$ scale factor, but $A(\eta)$ initiate the collapsing process, appear a bouncing and after acquires an increase mode. \label{111}}
				\end{center}
\end{figure}

In the Fig.(\ref{111}) we display the evolution for both scale factors.  We have the collapse of the scale factor $B(\eta)$, and the scale factor $A(\eta)$ have a subtle bouncing, and after a regret of the collapse process adopts a growing mode.\\
\\
$\bullet$ {\it \underline{Case with positive curvature}}\\
For the closed case Eq.(\ref{geral}) reads:
\begin{equation}
B(\eta) = B_\ast \left \lbrace \sin (\frac{\eta_{BC}- \eta}{2})   \right \rbrace ^2 \, ,
\end{equation}
where $B_\ast = \frac{B_i}{\sin (\eta_{BC}/2)}$. For $B_\ast = B_i \rightarrow \eta _{BC} = \Pi \, .$

\begin{figure}[pb]
	\begin{center}
	\includegraphics[scale=0.4]{scale-factor-A-kpositivo.pdf}
     \includegraphics[scale=0.4]{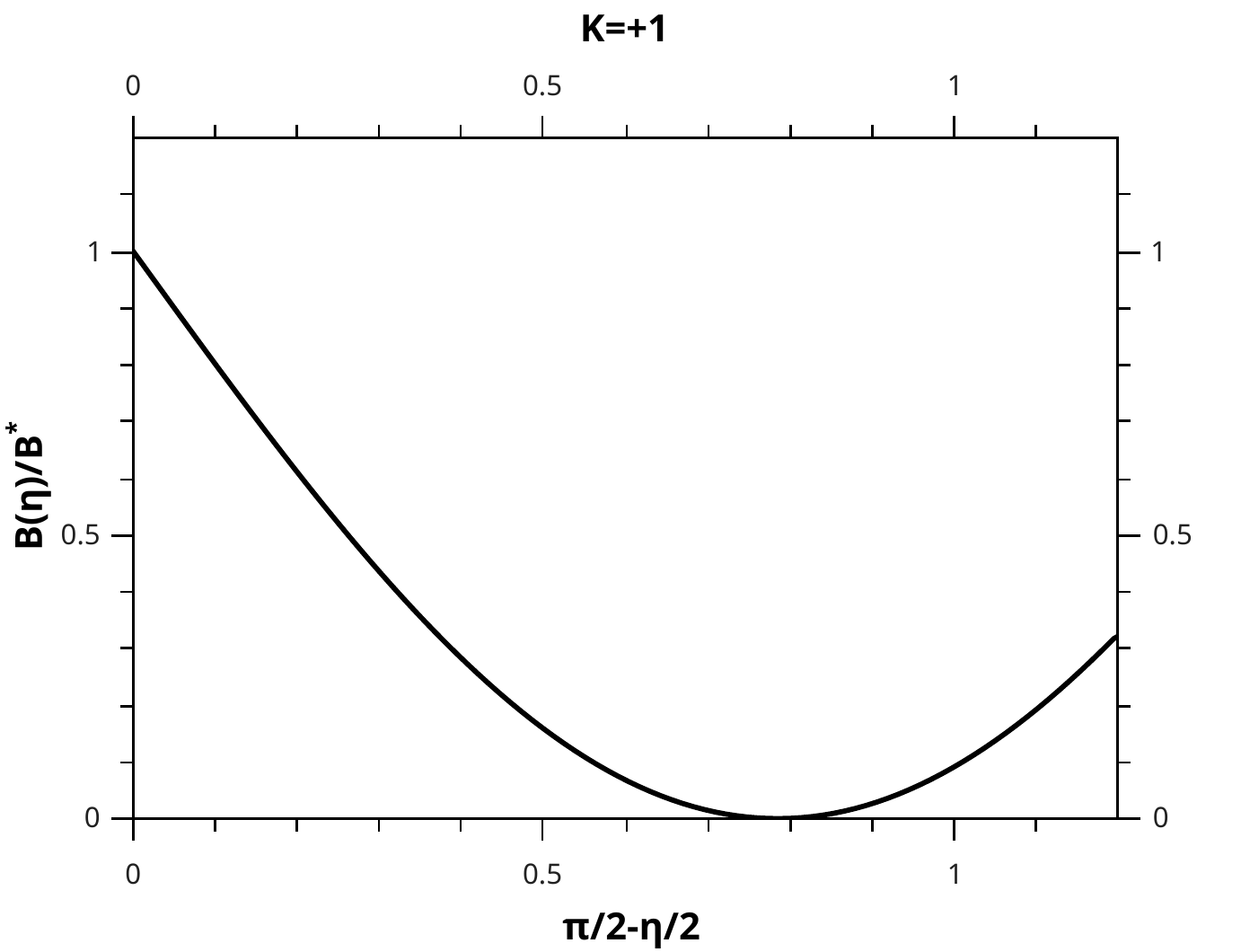}
	\caption{Evolution of the scale factor $A(\eta)$ and $B(\eta)$ for a positive curvature.
		In this case scale factor $B(\eta)$ collapses, but the scale factor $A(\eta)$ increases.  The singularity formation will, inevitably, not be punctual. \label{222}} 
\end{center}
\end{figure}

In turn, Eq.(\ref{A-eta}) assumes the form
\begin{equation}
A(\eta)=A_\ast \left \lbrace 2+\frac{(x-1)\sin x}{1+\cos x}\right \rbrace \, .
\end{equation}
In the Fig.(\ref{222}) we display the scale factors for this case.
Perhaps there is some more fundamental physical correlation, however, by looking at the process of analytic integration that we use in  field equations, the collapse of the spherical part of the line element implies survival of the radial dimension, which is a feature that we observe not only for the closed case, but for all other cases of curvature as well.
\\
$\bullet$ {\it \underline{Case with negative curvature}} \\
Finally, let's deal with the open case, which the expressions for the scale factors are:
\begin{eqnarray}
B(\eta)&\propto& (\sinh{\frac{1-\eta/\eta_{Bc}}{2}})^2 \, , \\
A(\eta) &\propto& -2+(2-\eta/\eta_{Bc})\frac{1+\cosh(1-\eta/\eta_{Bc})}{{\sinh(1-\eta/\eta_{Bc})}} \, ,
\end{eqnarray}
which the graphs are displayed in the Fig.(3).
\begin{figure}[tb]\label{333}
	\begin{center}
		\includegraphics[scale=0.4]{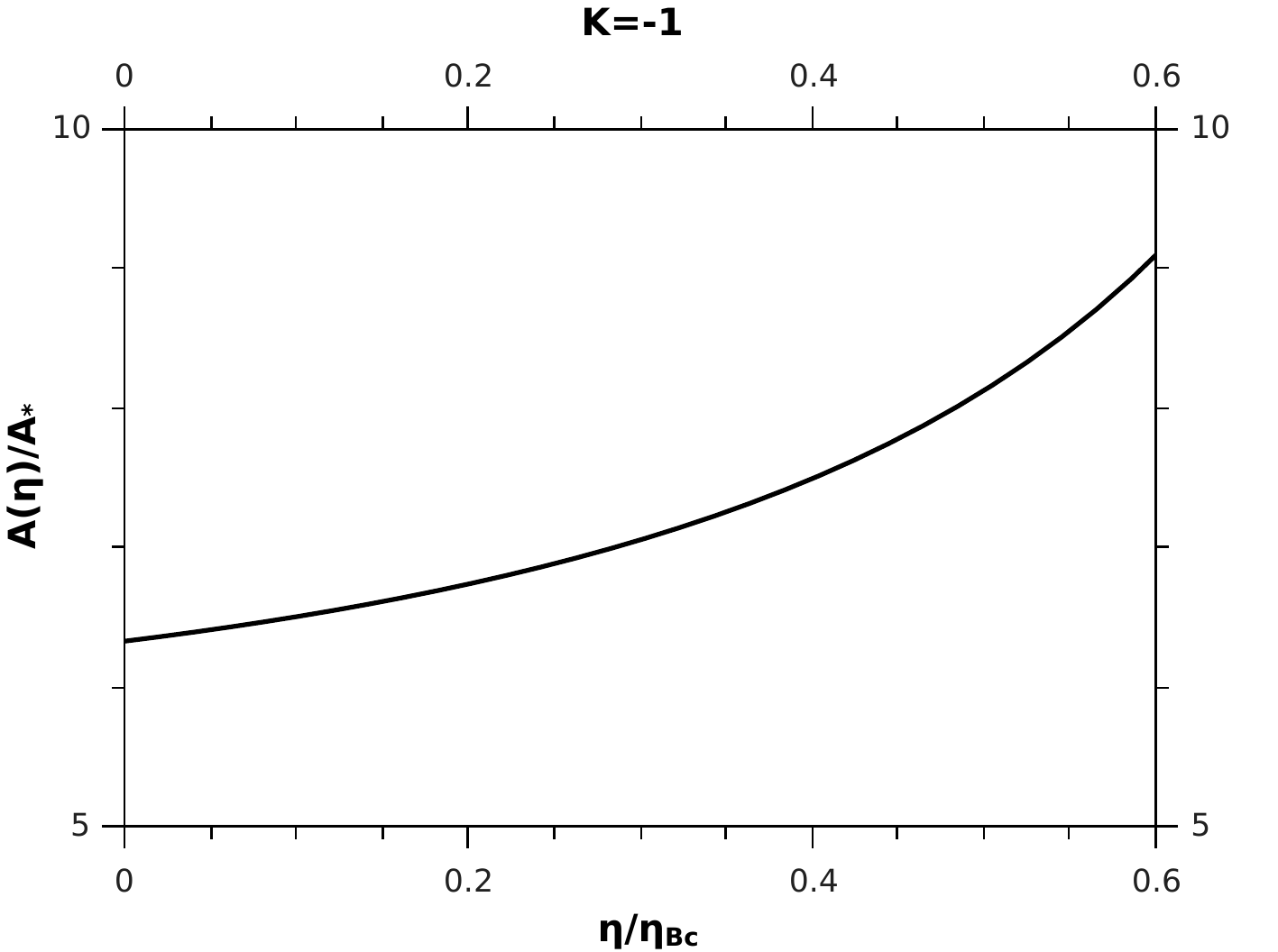} 
		\includegraphics[scale=0.4]{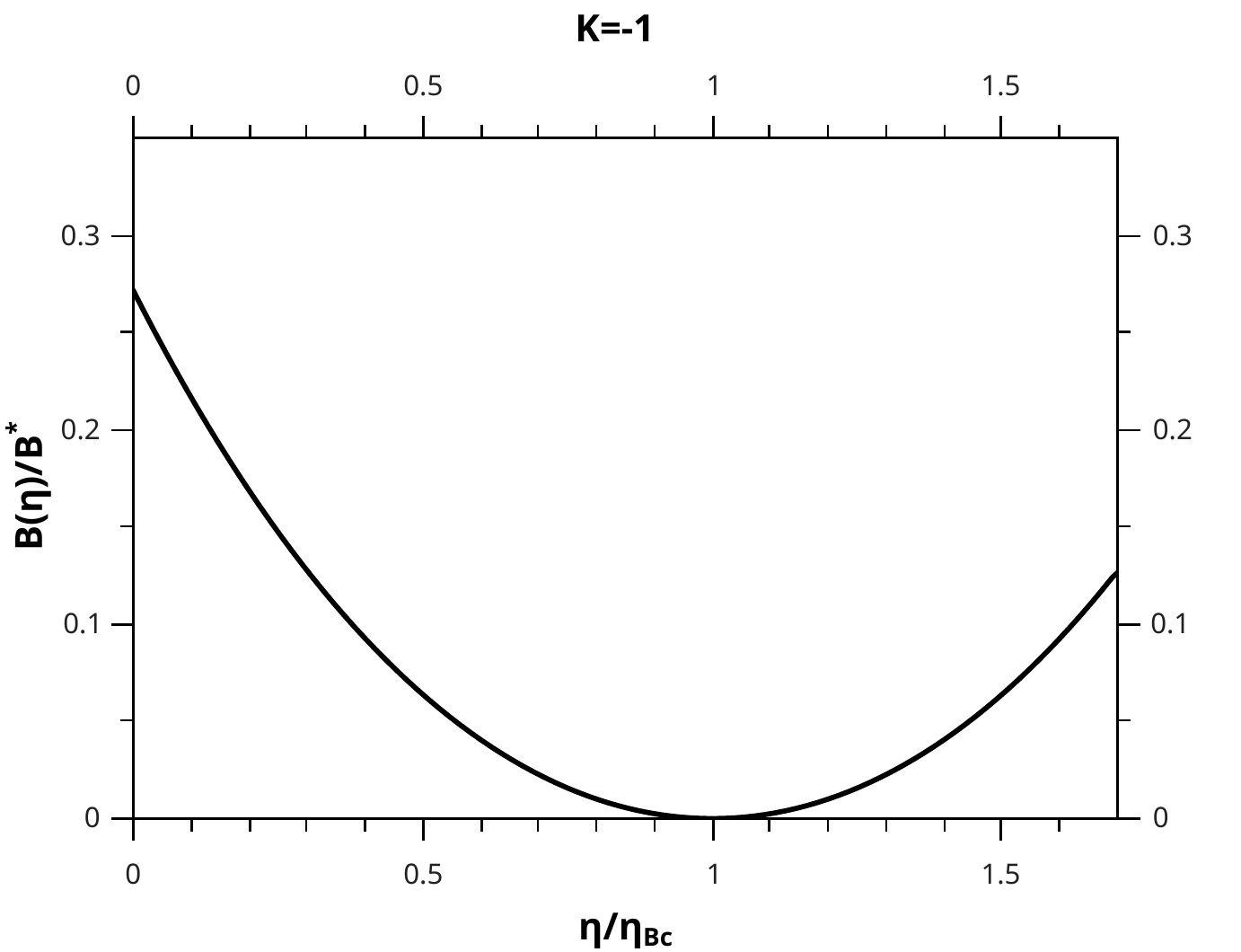} 
		\caption{Evolution of the scale factor $A(\eta)$ and $B(\eta)$, considering a negative curvature.} 
	\end{center}
\end{figure}
\section{Some features and constraints of the model}
Before moving on to the study of the gravitational collapse for KS model, we would like to consider some important features and constraints that have emerged in the scientific literature and we think it is appropriate to make some comments.
 
We have paved the way for an analytical and exact integration of field equations in our work. However, another means to the integration process is to consider an ansatz that be physically reasonable. For the KS model one can consider for example $\rho \propto \frac{1}{AB^2}$ as an ansatz for the energy density.  We note that, using the expressions for the scaling factors that we find for each curvature case and calculating the energy density: first using the ansatz given by equation $\rho \propto \frac{1}{AB^2}$; and then using the field equation for the energy density (Eq.\ref{fe1}), the expressions obtained retain a proportionality with each other, albeit with different constants depending on the curvature of spacetime.
 
Fundamentally, the search for additional links between certain quantities aims to reduce the relationship between the number of variables to be determined and the number of available equations. Following Ruslan and Tatiana, the authors defends the condition $\frac{\sigma}{H} = constant \, ,$ where $\sigma $ is the shear scalar and $H$ is the Hubble function.  They based this condition considering a perfect fluid with an anisotropic component for dark energy, and that $\frac{\sigma}{H} \approx 10^{-4}$ is compatible to the cosmic microwave background at the decoupling time \cite{Ruslan}.  Although this condition has been considered by the authors in a study of the cosmological point of view, in principle nothing prevents us from using it.

In the Fig.(4) we show the evolution of the derivative of quotient between the shear and the expansion, substituting the respective expressions for the scale factors.
\begin{figure}[tb]\label{444}
	\begin{center}
		\includegraphics[scale=0.7]{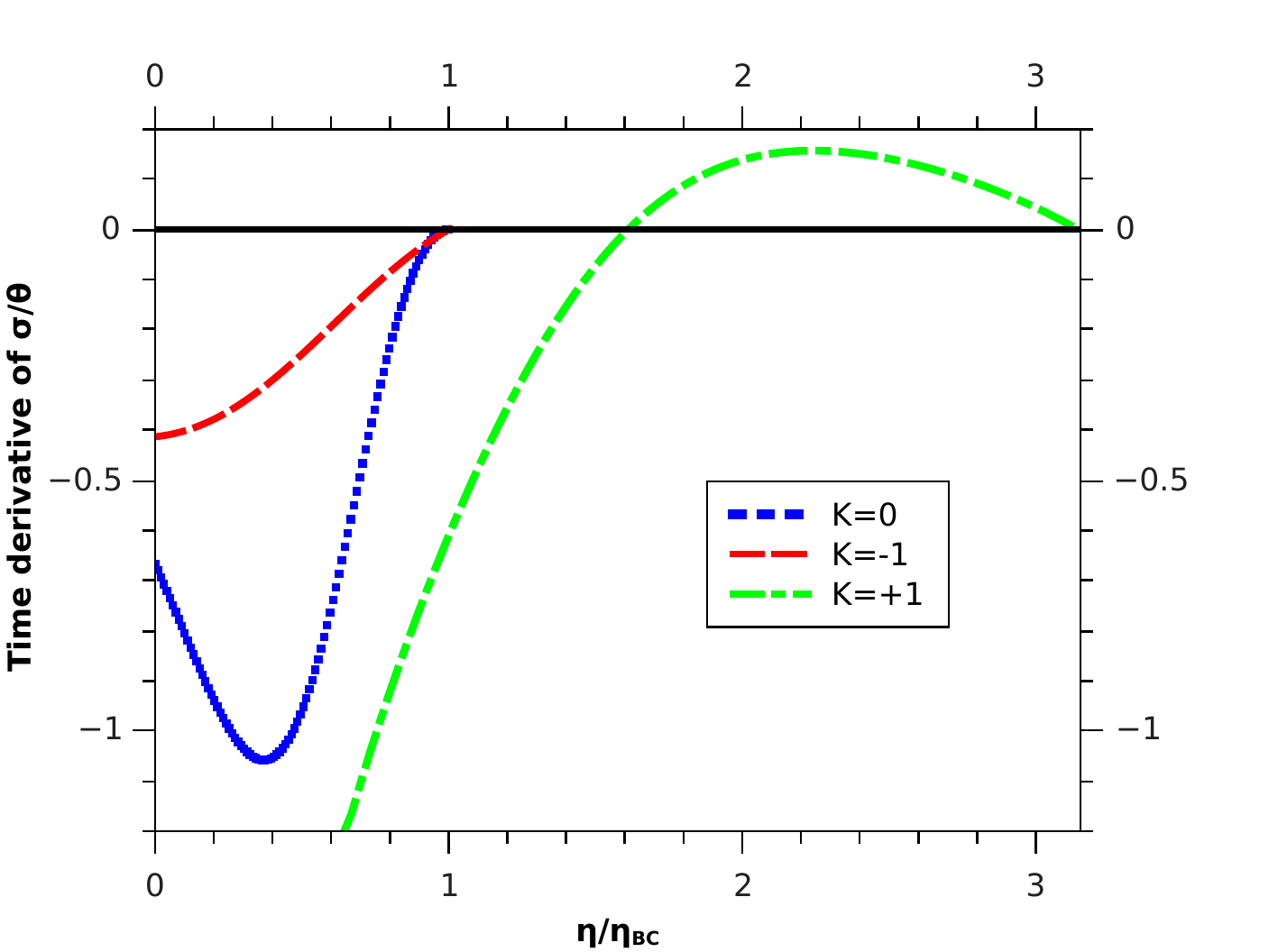} 
		\caption{Evolution  for the time derivative of quotient between the shear and Hubble function.} 
	\end{center}
\end{figure}
We can see that during the collapse process the derivative of the quotient between these quantities is only null at a specific time, for cases of zero or negative curvature. In the case of positive curvature, we will not have the nullity of the derivative during the collapse process.

Hence, with relative certainty that while the first discussed ansatz, involving the energy density, presented a very close characteristic of exact integration; the second case observed involving the ratio of shear to expansion does not seem to us to be a valid option as an ansatz.

Still on the shear, for the considered spacetime the shear tensor components are given by:
\begin{equation}
\sigma_{rr}=2/3\left\lbrace   \frac{A^\prime}{AB}-\frac{B^\prime}{B^2}   \right\rbrace =-2\sigma_{\theta \theta} = -2\sigma_{\phi \phi} \, .
\end{equation}
Naturally, if all shear components are null, results $\frac{A^\prime}{A} = \frac{B^\prime}{B}$, moving the model towards homogeneity and isotropy.
Assuming that the radial velocity of the fluid depends solely on the temporal coordinate, and taking into account the radial component of the shear, 
we can write that
$$\nabla \cdot \vec{v} \ \propto (\frac{A^\prime}{AB}-\frac{B^\prime}{B^2}) \, ,$$
where we have in the right side of equation the velocity divergence of the fluid.

For this, we take into account the expression for the radial composite of the shear stress tensor in spherical coordinates in the Newtonian framework.
We think that, in spite of consider the Newtonian framework the physical consequences of our expressions are not affected. For the three cases considering different curvatures the shear is an increase function, pointing to the increase of the gradient velocity, and in accord with the increase of the energy density. 

 In general the velocity gradient can be decomposed in rotation and strain tensors, and may contain important information about turbulence, however a supplementary decomposition can be used to refine the knownledge of flow structure and  dissipative process \cite{Keylock}.  Generally, within the framework of general relativity theory, a dissipative process is included via the creation of particles in the fluid; or by adding a term involving the vacuum energy of all fields that permeate the fluid; or even considering a viscous fluid.   
\section{Singularity formation, gravitational waves}
\begin{figure}[pb]
\begin{center}
	\includegraphics[scale=0.7]{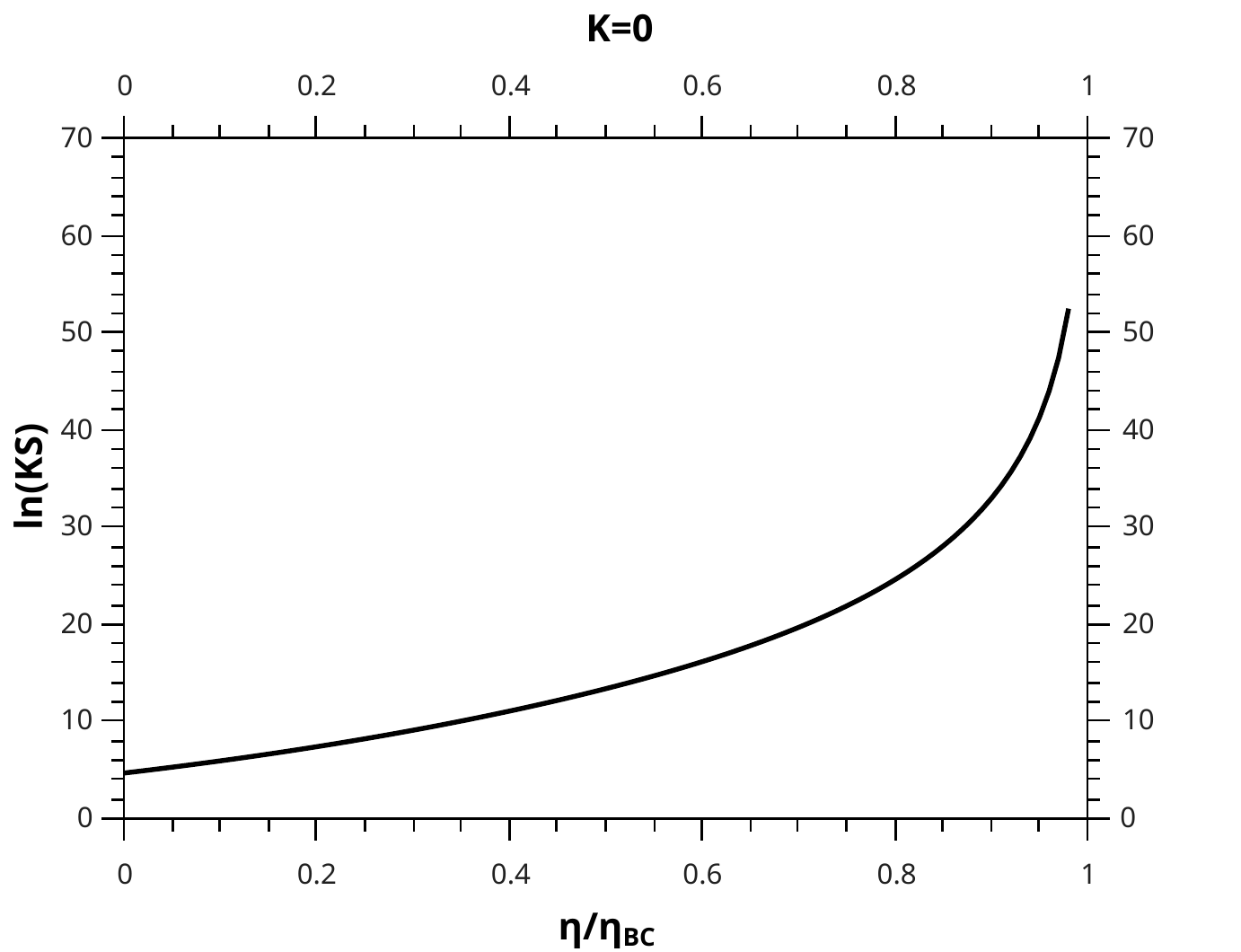}
	\caption{Evolution for the Kretschmann scalar (denoted in the graph by KS) for the null curvature.  For the other cases we have
		a similar behavior. Consequently, we can said that we have truth singularities for the all cases of curvature, since the Kretschmann scalar diverges. \label{kk}}
		\end{center}
\end{figure}

Taking into account the behavior of the Kretschmann scalar, at first, furnish to us a better characterization
of the singularity formation, after simple but horrendous calculations \cite{Richard}.  Hence, for KS spacetime the Kretschmann  scalar is given by:
\begin{equation}\label{Krets}
K=\frac{4}{B^4}\left\lbrace \frac{\ddot{A}^2}{A^2}-2\frac{\ddot{A}\dot{A}\dot{B}}{A^2B}+3\frac{\dot{B}^2 \dot{A}^2}{A^2B^2}+2\frac{\ddot{B}^2}{B^2}
-4\frac{\ddot{B}\dot{B}^2}{B^3}+3\frac{\dot{B}^4}{B^4}+K^2-2\frac{K\dot{B}^2}{B^2} \right\rbrace \, .
\end{equation}
Although at present time this calculation is not horrendous if you use one of the algebraic computing software. However, care must be taken with which version to use since that the errors committed by the tensorial packages have been fixed over time.
To analyze the obtained expressions for the Kretschmann scalar, substituting the scale factors
for each curvature case results in very extensive expressions.  A more immediate analyze can be realized using the respective
graphs. As the case with flat spacetime reproduces a behavior similar to that of more cases, we restrict ourselves to only showing the graph for $K=0$, which are displayed in the Fig.(\ref{kk}).  Now, it is easy to claim that we have  emerging singularities for all cases with different curvatures.
Although some more information can be taken from the Kretschmann scalar, it does not present a conclusion regarding the final fate of the collapse process.

As the radial dimension of the spacetime during collapse process survives, it is interesting to note that at some point our model in question was directed towards the collapse of the spherical part of spacetime, probably when we consider the fluid to be perfect.  However, we think this point deserves a closer look in the future.

We can characterize the object that is formed during the process of gravitational collapse, observing the appearance of the apparent horizon, which  are spacelike surfaces with future point convergence null geodesics on the both sides of surfaces \cite{Hawking}. If the moment of formation of the apparent horizon precedes the moment of formation of the singularity we will have a black hole, in case of a point collapse. Otherwise the singularity is called naked.  

For  the metric space given by Eq.(\ref{metric}) the formation of the apparent horizon, governed by 
$$\tilde{r}_{B ,\alpha} \tilde{r}_{B,\beta} g^{\alpha \beta} = 0 $$
 reduces to equation \cite{Campos}
\begin{equation}\label{primer}
rB^\prime = \frac{B^2}{A} \, ,
\end{equation}
where the areal radius is $\tilde{r}_B = rB(\eta=\eta_{AH})$. On the other hand, we can write Eq.(\ref{primer})
as
\begin{equation}\label{new}
\tilde{r}_A H_B = \pm 1 \, ,
\end{equation}
with $\tilde{r}_A = rA(\eta = \eta_{AH})$.
While the positive signal is relative to the expansion process, the negative signal characterize the collapsing fluid.
Considering $rA_\ast$ as unity, the equation that decides if we have a naked singularity, or a black object, assumes the form
\begin{equation}\label{new1}
H_B |_{\eta=\eta_{AH}} = -\frac{A_\ast}{A}\, .
\end{equation}
We will analyze each case of curvature on an individual basis.

So,considering the null curvature case we have
\begin{equation}\label{initial}
\frac{(1-\eta_{AH}/{\eta_{BC}})^3}{(1-\eta_{AH}/\eta_{BC})^{-1}+(1-\eta_{AH}/{\eta_{BC}})^2} = |H_{Bi}|\, .
\end{equation} 
The apparent horizon emerge before the system reach the singularity, and we have the formation of a black pencil.
 
In respect to the positive curvature case, we use the respective expressions for $A(\eta)$ and $B(\eta)$, substituting in Eq.(\ref{new1}) and we obtain the 
root numerically, resulting $\eta \approx 0.5465$.  Hence, any value for the the conform time $\eta$ anterior to this value privileges the formation of a black pencil.
Notice, that the collapsing time for this case is $\eta_{Bc} = \Pi$. We will do another brief comment, now for pedagogical point of view: for $\eta > 0.5465$,  the Eq.(\ref{initial}) is not longer valid, since the left side of equation is a negative quantity.

Finally, considering the negative curvature case we have a different situation, using identical procedure that we use for the closed case, making use of numerical procedure for the determination of the root of the Eq.(\ref{new1}), the final fate of the collapsing process is a naked pencil.  Since that, we does not obtain a value for the apparent horizon moment before the reaching of the singularity.

Now, in respect to the gravitational waves formation due to the collapse process, we can an idea about the formation of them
using the Weyl scalars, which for KS spacetime are:
\begin{eqnarray}
\Psi_1 &=& 0 \\
\Psi_3  &=& 0 \\
\Psi_2 &=& \frac{1}{12B^2}\left \lbrace  -\frac{\ddot{A}}{A} +2\frac{\dot{B}\dot{A}}{AB}-2\frac{\dot{B}^2}{B^2}+\frac{\ddot{B}}{B}+\frac{f^{\prime \prime}}{f} \right\rbrace\\
\Psi_4 &=& \frac{1}{8 f^2 B^4}\left \lbrace  -\frac{\ddot{A}}{A} +2\frac{\dot{B}\dot{A}}{AB}-2\frac{\dot{B}^2}{B^2}+\frac{\ddot{B}}{B}+\frac{f^{\prime \prime}}{f} \right\rbrace\\
\Psi_0 &=& \frac{f^2}{2} \left \lbrace -\frac{\ddot{A}}{A}+2\frac{\dot{A}\dot{B}}{AB}-2\frac{\dot{B}^2}{B^2}+\frac{\ddot{B}}{B} -\frac{f^{\prime \prime}}{f}\ \right\rbrace \, .
\end{eqnarray}
Since we have $\Psi_1 = \Psi_3 = 0$, our spacetime admits null directions with multiplicity 2, that is, 
we have the existence of two double principal null directions, and consequently, we have a type D in Petrov classification.
\section{Some comments and conclusions}
Some time ago, Assad and Lima \cite{Assad}, find a general and unified solution for spatially homogeneous and isotropic cosmologies
containing a perfect fluid in the material part of the field equations.  The same procedure was used to study   the gravitaional collapse process in a
Friedmann geometry with arbitrary curvature \cite{Perico}.  In this work, we used the original method of Assad and Lima to integrate the field equations for the Kantowski-Sachs model considering an arbitrary curvature.  We find exact and explicit solutions for the field equations, which are studied
in accord with the gravitational collapsing process, and the singularities are characterize using the Kretschann scalar.

Although the Kretschmann scalar is useful in elucidating the formation of a real singularity, and not due to the particular coordinate system, nothing at all tells us about the characteristics of the object formed at the end of the gravitational collapsing process. However, considering the apparent horizon formation moment we have as the final state for the collapsing process a black, or a naked pencil.  The first case for the positive and null curvatures, and the second for the negative curvature case.

With respect to the components of the shear tensor, if there were an evolution to a null value, we would have as consequence the evolution of energy density to a constant value, and this does not happen.

Finally, we determine the Petrov classification and notice the existence of gravitational waves, since that we have an type D spacetime in Petrov classification. 

Recently, a work appeared concerning to the light redshift for an anisotropic geometry, where the authors conclude that it depends
of propagation direction and it may become dispersive \cite{Sergio}.  This is a point that I would like to devote some attention to in the future involving KS geometry, as well as the decomposition of the velocity gradient and its relationship to the structure of the stellar fluid flow and dissipation process.



\begin{thebibliography}{0}
\bibitem{Mendez} Luis. E. Mendez and Alfredo B. Henriquez, Physics Letters B {\bf 254} (1991) 44.
\bibitem{Roy} Subenoy Chakraborty and Anusua Roy, Astrophys Space Sci. {\bf 313} (2008) 389.
\bibitem{Adhav} Kishor S. Adhav, Abhijit S. Bansod, Rajesh P. Wankhade and G. Ajmire Hanumant, Cent. Eur. J. Phys. {\bf 9} (2011) 919.
\bibitem{Skyrme} Luca Parisi, Ninfa Radicella and Gaetano Vilasi, Phys. Rev. D {\bf 91} (2015) 063533. 
\bibitem{Katore} S. D. Katore and S. P. Hatkar, Prog. Theor. Exp. Phys. {\bf 2016} (2016) 033E01.
\bibitem{Saridarkis} Genly Leon and Emmanuel N. Saridakis, Class.Quant.Grav. {\bf 28} (2011) 065008.
\bibitem{Ray} Pratik P. Ray, B. Mishra and S. K. Tripathy, International Journal of Modern Physics D {\bf 28} (2019) 1950003.
\bibitem{Abbas} G. Abbas, Astrophys. Space Sci. {\bf 252} (2014) 955.
\bibitem{Maurya} S. K. Maurya and Y. K. Gupta and Saibal Ray and Baiju Dayanandan, Eur. Phys. J. C. {\bf 75} (2015) 225.
\bibitem{Bradley} Michael Bradley, Peter K S Dunsby, Mats Forsberg and Zoltán Keresztes, Classical and Quantum Gravity {\bf 29} (2012) 095023. 
\bibitem{Dorosh} A. G. Doroshkevich, Astrofizika {\bf 1} (1965) 255.
\bibitem{Abbo} B. P. Abbot et al, arXiv:1711.05578 (2017).
\bibitem{Kantowski} R. Kantowski and R. K. Sachs, J. Math. Phys. {\bf 7}, (1966) 443.
\bibitem{Ellis} George F. R. Ellis, Roy Marteens and Malcolm A. H. MacCallum, {\it Relativistic Cosmology}, Cambridge University Press (2012); 
\bibitem{Campos} M. Campos and J. A. S. Lima,  Phy. Rev. D {\bf 86}, 043012 (2012), arXiv:1207.5150 [gr-qc].
\bibitem{Perico} E. L. D. Perico, J. A. S. Lima and M. Campos, Gen. Rel. Grav. {\bf 48}, (2016) 8. 
\bibitem{Ruslan} Ruslan K. Muharlyamov, Tatiana N. Pankratyeva, Astrophys. Space Sci. {\bf 363}, (2018) 32.
\bibitem{Keylock} Cristopher J. Keylock, Jornal of  fluid mechanics {\bf 848}, (2018) 876.
\bibitem{Richard} Richard Conn Henry, (1999) arXiv:9912320v1 [astro-ph]
\bibitem{Hawking} S. W. Hawking and G. F. R. Ellis, {\em The Large Scale Structure of Spacetime}, Cambridge University Press, Cambridge  (1973).
\bibitem{Assad} M. J. D. Assad and J. A. S. Lima, Gen. Rel. Grav. {\bf 20}, (1988) 527.
\bibitem{Sergio} Sergio A. Hojman and Felipe A. Araujo, arXiv:1801.05472v1 [gr-qc].
\end{thebibliography}


\end{document}